\documentclass[final,english]{bullsrsl}[2022/06/15]

\usepackage[latin1]{inputenc}
\usepackage[T1]{fontenc}

\usepackage{natbib} 
\usepackage{graphicx}
 
 \usepackage{amsmath}
 \usepackage{epstopdf,epsfig}
 \usepackage{epsf}
 \usepackage{latexsym}
 \usepackage{amssymb}
 \usepackage{longtable}
 \usepackage{float}

\begin{document}
\title{Prolonged and Extremely Non-radial Solar Wind Flows}

\author[affil={1}, corresponding]{Susanta K.}{Bisoi}
\author[affil={2}]{Diptiranjan}{Rout}
\author[affil={3}]{P.}{Janardhan}
\author[affil={4}]{K.}{Fujiki}
\author[affil={5}]{Dibyendu}{Chakrabarty}
\author[affil={1}]{Karan}{Sahu}
\affiliation[1]{Department of Physics and Astronomy, National Institute of Technology, Rourkela -769008, India}
\affiliation[2]{National Atmospheric Research Laboratory, Gadanki, India}
\affiliation[3]{Astronomy \& Astrophysics Division, Physical Research Laboratory, Navrangpura, Ahmedabad, India.}
\affiliation[4]{Institute for Space-Earth Environmental Research, Nagoya, Japan}
\affiliation[5]{Space \& Atmospheric Science Division, Physical Research Laboratory, Navrangpura, Ahmedabad, India.}
\correspondance{bisois@nitrkl.ac.in}
\date{07 July 2023}
\maketitle

% Abstract of the paper in the same language as the paper
\begin{abstract}
 We present a study of three highly non-radial solar wind events when the azimuthal solar wind flow angle exceeds $>$ 6${^{\circ}}$ for one day or more. None of the events are associated with coronal mass ejections and co-rotating interaction regions observed at 1 AU. For all events, the solar wind outflows at 1 AU have low solar wind velocity and solar wind density. Based on the significant increase in the Oxygen charge state ratio of ${O^{7+}}/{O^{6+}}$ at 1 AU for all of the events, we have traced them back to the Sun and found that their source regions originated in an active region and coronal hole (AR-CH) pairs mainly located at the central meridian. Further, examining the dynamical evolutions in their source regions using both the {{Extreme ultra-violet Imaging Telescope}} and {{Michelson Doppler Imager}}, it is found that the changes taking place in AR-CH boundaries eventually disturbed the stable CH configurations, resulting in a reduction of the CH area and finally its disappearance, leaving only with the AR. Our study provides a possible explanation to discuss the origin of the prolonged and highly non-radial solar wind flows.
\end{abstract}


\section{Introduction}
The solar corona's high temperature (10$^{6}$ K) helps coronal plasma escape the Sun's gravitational potential well and flow upward along open magnetic field lines to form the solar wind. The solar wind drags the frozen-in magnetic field out to 1 AU and beyond in the well-known Parker's spiral, and therefore, it is generally radial beyond the Alfv\'en radius (R${_{A}}$). However, it is observed on several occasions that the solar wind flows have been highly non-radial where the azimuthal component of the solar wind velocity (V$_{y}$) is going well above the usual range 10\,--\,30 km/s or the azimuthal solar wind flow angle ($\Phi_{v}$ = arctan($\frac{-V_{y}}{V_{x}}$)) is more than $6^{\circ}$. The X- and Y-component of solar wind velocity are in the Geocentric Solar Ecliptic (GSE) coordinate system. In the GSE coordinate system, the X-axis points towards the Sun, the Y-axis is in the ecliptic plane, and points towards the dusk, and the Z-axis is parallel to the ecliptic pole along the positive north direction.
Primarily, high non-radial flows have been observed when the solar wind flows were associated with coronal mass ejections (CMEs) or their interplanetary counterparts, {\it{i.e.}}, interplanetary coronal mass ejections (ICMEs) where the azimuthal component of solar wind velocity can be significantly higher  ($\ge 100$ km/s)\citep{OCa04}.  Generally, CMEs/ICMEs are fast-moving and contain a well-defined magnetic cloud. The magnetic field structures within the ICME and its embedded magnetic cloud are observed to be anchored to the Sun's surface even when they arrive at 1 AU. The highly non-radial flows due to CMEs occur because of the deflection of the solar wind plasma around the draped magnetic field structures. The draping of the magnetic field structures occurs around the CME plasmoid because of the low beta plasma condition, which arises due to the expansion of the CME cloud in space. Besides CMEs, highly non-radial flows have also been observed for co-rotating solar wind structures, such as co-rotating interaction regions (CIRs) \citep{GoP99}.  A stream interaction region (SIR) usually forms when the faster solar wind originating from the coronal holes (CH) overtakes the background slower solar wind \citep{JiR06, RiC18}. They are classified as CIRs when they exist for at least one complete solar rotation \citep{JiR06}.  The compression of the faster and slower solar wind streams makes the solar wind flow non-radial for a few hours during the passage of CIRs.

Highly non-radial flows, lasting over 24 hours, have also been observed in the absence of CMEs/CIRs, known as ``solar wind disappearance events'' \citep{JaF08, JaT08}. During these events, other than highly non-radial flows, low density (\textless~0.1 cm$^{-3}$), low velocity (\textless~350 kms$^{-1}$) solar wind outflows associated with large flux expansion factors, and extended Alfv\'en radii are observed \citep{JaF05}. The solar sources of these unique events are found to be small, short-lived (24\,--\,48 hours), transient mid-latitude CH \citep{KHu01}, located at the central meridian and adjacent to active regions (AR). \cite{JaT08} for such a solar wind disappearance event on 11 May 1999 suggested that the dynamical evolution of the AR and CH pinch-off the transient solar wind outflow coming from the CH to result in a low solar wind density at 1 AU. In the present study, a detailed investigation has been carried out on three extreme non-radial events where the solar wind flow angle exceeds 6${^{\circ}}$ for 24 hours or more, and no CMEs/CIRs are observed during all of the events. 
\section{Observations}
\subsection{The Events of 10--12 May 1999, 27--30 Mar 2000 and 15--18 Feb 2004}
\begin{figure}
\centering
\includegraphics[height=0.60\textwidth,width=.70\textwidth]{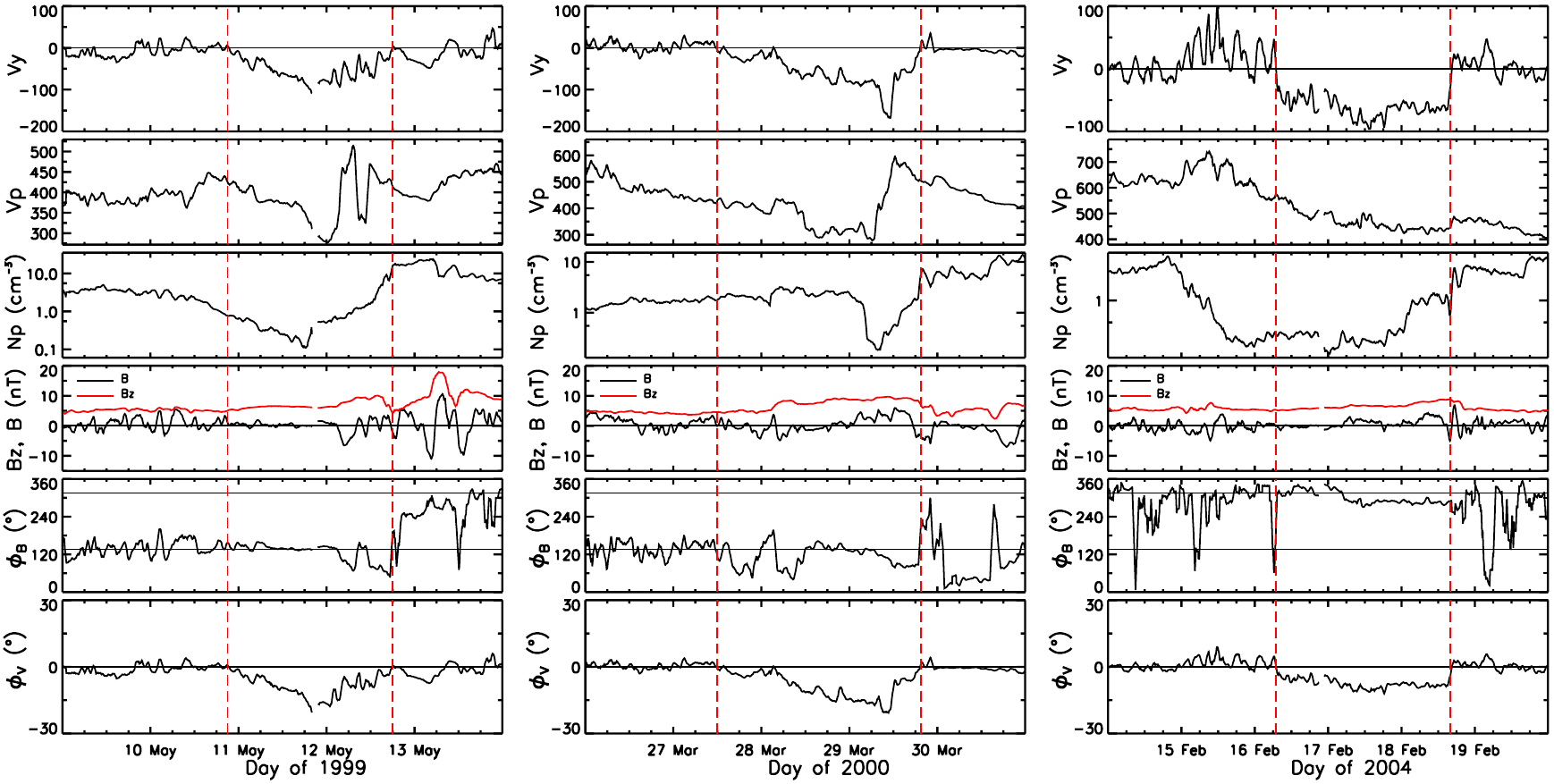}
\bigskip
\begin{minipage}{12cm}
\caption{Variations for solar wind parameters for the events of May 1999 (left), Mar 2000 (middle), and Feb 2004 (right). The parameters, from top to bottom, are the Y-component of solar wind velocity (V$_{y}$), the solar wind flow velocity (V$_{p}$), proton density (N$_{p}$), the Z-component of interplanetary magnetic field ($B_{z}$), the magnetic field strength ($\mid$B$\mid$), the solar wind azimuthal magnetic field angle ($\Phi_{B}$), and the solar wind azimuthal flow angle ($\Phi_{v}$), respectively. The vertical lines in red demarcate the duration of non-radial solar wind flows.}
\label{fig1}
\end{minipage}
\end{figure}

Figure \ref{fig1} presents observations of solar wind plasma and magnetic field parameters in the GSE coordinate system at the first Lagrangian point (L1) from the {\it{OMNI}} database as a function of the day during the events of May 1999 (left), Mar 2000 (middle) and Feb 2004 (right). In each panel, the solar wind plasma and magnetic field parameters, from top to bottom, are the Y-component of solar wind velocity (V$_{y}$), the solar wind flow velocity (V$_{p}$), proton density (N$_{p}$), the Z-component of interplanetary magnetic field ($B_{z}$), the magnetic field strength ($\mid$B$\mid$), the solar wind azimuthal magnetic field angle ($\Phi_{B}$), and the solar wind azimuthal flow angle ($\Phi_{v}$), respectively. As mentioned earlier, the large values of both V$_{y}$ and $\Phi_{v}$ indicate the non-radial flow nature of the solar wind.
It is evident from Fig. \ref{fig1} (left) that the $\Phi_{v}$ has been non-radial since $\sim$21:00 UT of 10 May, and after that, it has continued to be non-radial for the next 44 hours. The non-radial flow of the solar wind is also evident from the negative value of V$_{y}$, reaching as high as $\sim$125 kms$^{-1}$. The strength of the $V_{p}$ has also continuously decreased. Similarly, a continuous reduction in N$_{p}$ has been observed, remaining $\textless 1$ cm$^{-3}$ for more than 24 hours. On the other hand, the $B_{z}$ is mostly northward, and its magnitude always hovers around the zero line, while the value of ${\mid}B{\mid}$ has remained steady. 
%----------------------------- Begin Fig 1 -----------------------
%
% \begin{center}
% 	\protect\begin{figure*}[!htb]
% 	       \centering
% 		\includegraphics[height=0.75\textwidth,width=.9\textwidth]{Fig1.pdf}
% 		\caption{Variations for solar wind parameters for the events of May 1999 (left), Mar 2000 (middle), and Feb 2004 (right). The parameters, from top to bottom, are the Y-component of solar wind velocity (V$_{y}$), the solar wind flow velocity (V$_{p}$), proton density (N$_{p}$), the Z-component of interplanetary magnetic field ($B_{z}$), the magnetic field strength ($\mid$B$\mid$), the solar wind azimuthal magnetic field angle ($\Phi_{B}$), and the solar wind azimuthal flow angle ($\Phi_{v}$), respectively. The vertical lines in red demarcate the duration of non-radial solar wind flows.}
% 		\label{fig1} 
% 	\end{figure*}
% 	\vspace{-0.5cm}
% \end{center}
%----------------------------------- End Fig 1 ------------------------

The $\Phi_{v}$ is also negative since 13:00 UT of 27 Mar 2000, as evident from Fig. \ref{fig1} (middle), which has continued to be negative for the next 56 hours. Further, continuous negative variations of V$_{y}$ confirm the non-radial flow of solar wind during the Mar 2000 event. During the non-radial flow of solar wind, Vp values have been reduced. The value of N$_{p}$ has also remained $\textless 1$ cm$^{-3}$ for a continuous $\sim$18 hours. On the other hand, the $B_{z}$ is mostly northward, and the ${\mid}B{\mid}$ is nearly steady during the non-radial flow interval. Figure \ref{fig1} (right) shows another long-duration non-radial flow event in Feb 2004. It is evident from Fig. \ref{fig1} (right) that the solar wind flow is non-radial for around 58 hours, showing a negative value for $\Phi_{v}$ during the period 15\,--\,18 Feb 2004. A continuous negative value of Vy is seen during the period, further showing the solar wind's prolonged non-radial flow nature during 15\,--\,18 Feb 2004. In addition, the solar wind velocity shows continuous reductions. However, it has never gone below 400 kms$^{-1}$, and the $B_{z}$ mostly hovers around the zero line, like the earlier events. The non-radial solar wind flow also has a low density, less than $\textless 1$ cm$^{-3}$ for over two days. 
%----------------------------- Begin Fig 2-----------------------
%
% \begin{center}
% 	\protect\begin{figure}[!ht]
% 	\centering
% 		\includegraphics[height=0.48\textwidth, width=.48\textwidth]{Fig2.pdf}
% 		\caption{As a function of day, the charge state ratio ${O^{7+}}/{O^{6+}}$ plotted 
% 		for the period, top) 8\,--\,15 May of 1999, middle) 25 Mar\,--\,3 Apr of 2000, and bottom) 
% 		14\,--20\,Feb of 2004. The horizontal dashed line is drawn at ${O^{7+}}/{O^{6+}}$ 
% 		ratio of 0.2, while the vertical dashed lines in red demarcate the period corresponding to the 
% 		non-radial flow, as indicated in Fig. \ref{fig1}, during which the Oxygen charge state ratio is well above 0.2.}
% 		\label{fig2} 
% 	\end{figure}
% 	\vspace{-0.9cm}
% \end{center}
\begin{figure}
\centering
\includegraphics[scale=0.3]{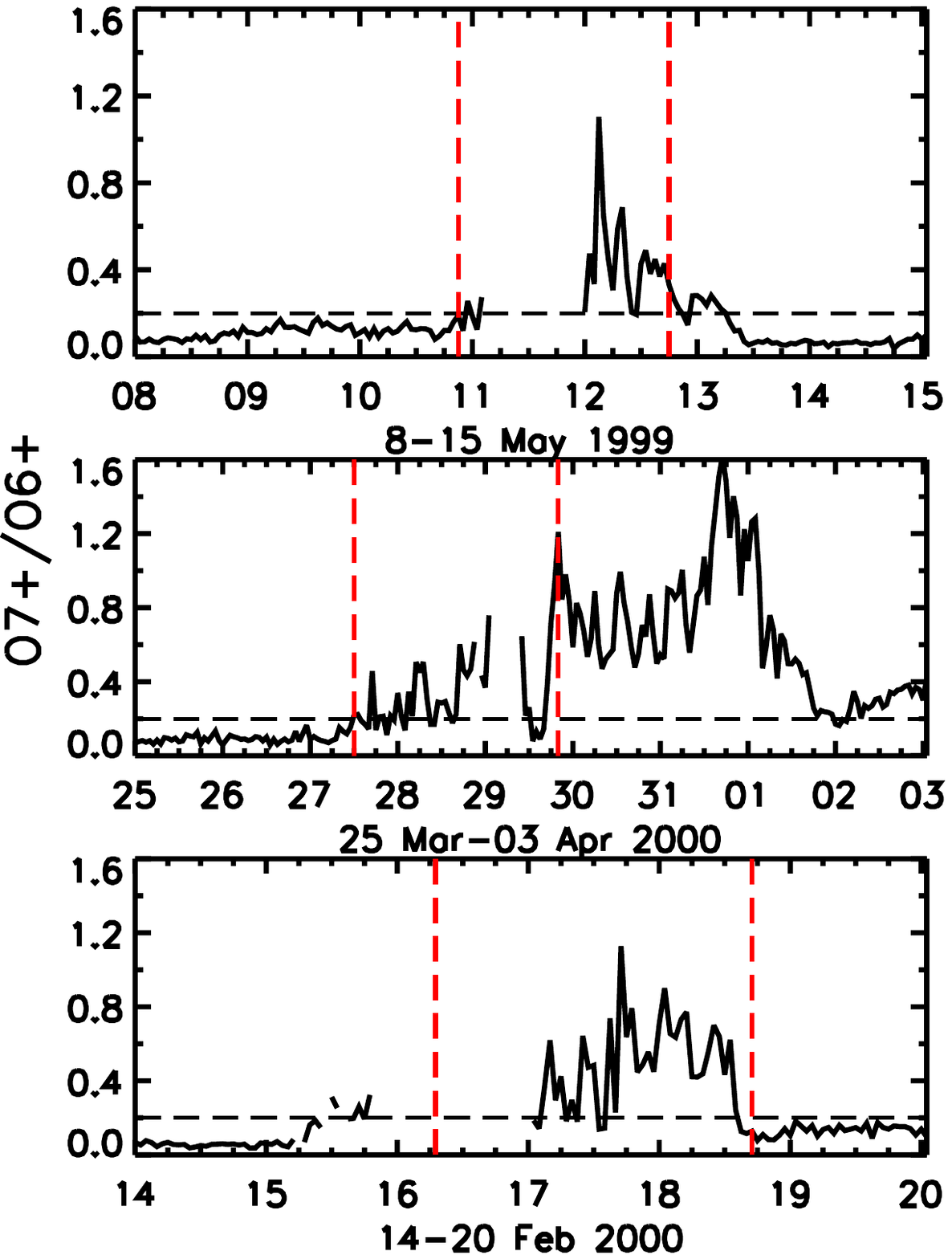}
\bigskip
\begin{minipage}{12cm}
\caption{As a function of day, the charge state ratio ${O^{7+}}/{O^{6+}}$ plotted 
 		for the period, top) 8\,--\,15 May of 1999, middle) 25 Mar\,--\,3 Apr of 2000, and bottom) 14\,--20\,Feb of 2004. The horizontal dashed line is drawn at ${O^{7+}}/{O^{6+}}$ 
 		ratio of 0.2, while the vertical dashed lines in red demarcate the period corresponding to the non-radial flow, as indicated in Fig. \ref{fig1}, during which the Oxygen charge state ratio is well above 0.2.}
\label{fig2}
\end{minipage}
\end{figure}
%----------------------------------- End Fig2 ------------------------
The charge state ratio of Oxygen (${O^{7+}}/{O^{6+}}$) measured at L1 by {{Advanced Composition Explorer}} ({{ACE}}) satellite is plotted in Fig. \ref{fig2} for the events of May 1999 (top), Mar 2000 (middle), and Feb 2004 (bottom). The solar wind from AR sources usually has a higher (0.1\,--\,0.6) ${O^{7+}}/{O^{6+}}$ ratio than from CH (0.2) regions \citep{LiP04}. It is evident from Fig. \ref{fig2} that the Oxygen charge state ratio ${O^{7+}}/{O^{6+}}$ is below 0.2 and has started increasing for all the events at the time when the solar wind flow becomes highly non-radial. It is to be noted that the Oxygen charge state ratio is well above 0.2 during the whole duration of non-radial flow, as demarcated by the vertical lines in red in each panel of Fig. \ref{fig2}. The ${O^{7+}}/{O^{6+}}$ ratio $>$ 0.2 indicates a clear transition of solar wind source regions from CH to AR. 
%
%
%----------------------------- Begin Fig 3 -----------------------
 %
% \begin{center}
%  	\protect\begin{figure*}[htb]
%  		\centering
%  		\includegraphics[height=0.68\textwidth,width=.58\textwidth]{Fig3.pdf}
%  		\caption{Synoptic maps, made using {{SOHO/MDI}} magnetograms, during CR1949 (top),  CR1961 (middle), and CR 2013 (bottom) corresponding to May 1999, Mar 2000, and Feb 2004 events. Regions of strong magnetic fields corresponding to active region locations are shown as black and white patches representing negative and positive magnetic polarities, respectively. The solid curved lines in yellow on each map are the source surface magnetic neutral lines. The groups of converging lines in red and magenta colors are potential field computed \citep{HKo99} magnetic field lines that join the source surface at 2.5 R$_{\odot}$ to their counterparts at the photosphere. The lines in red in each panel particularly indicate the footpoints of the converging field lines during the traceback period when the solar wind flow was highly non-radial.}
%  			\label{fig3} 
%  	\end{figure*}
% 	\vspace{-0.8cm}
%  \end{center}
\begin{figure}[!ht]
\centering
\includegraphics[scale=0.5]{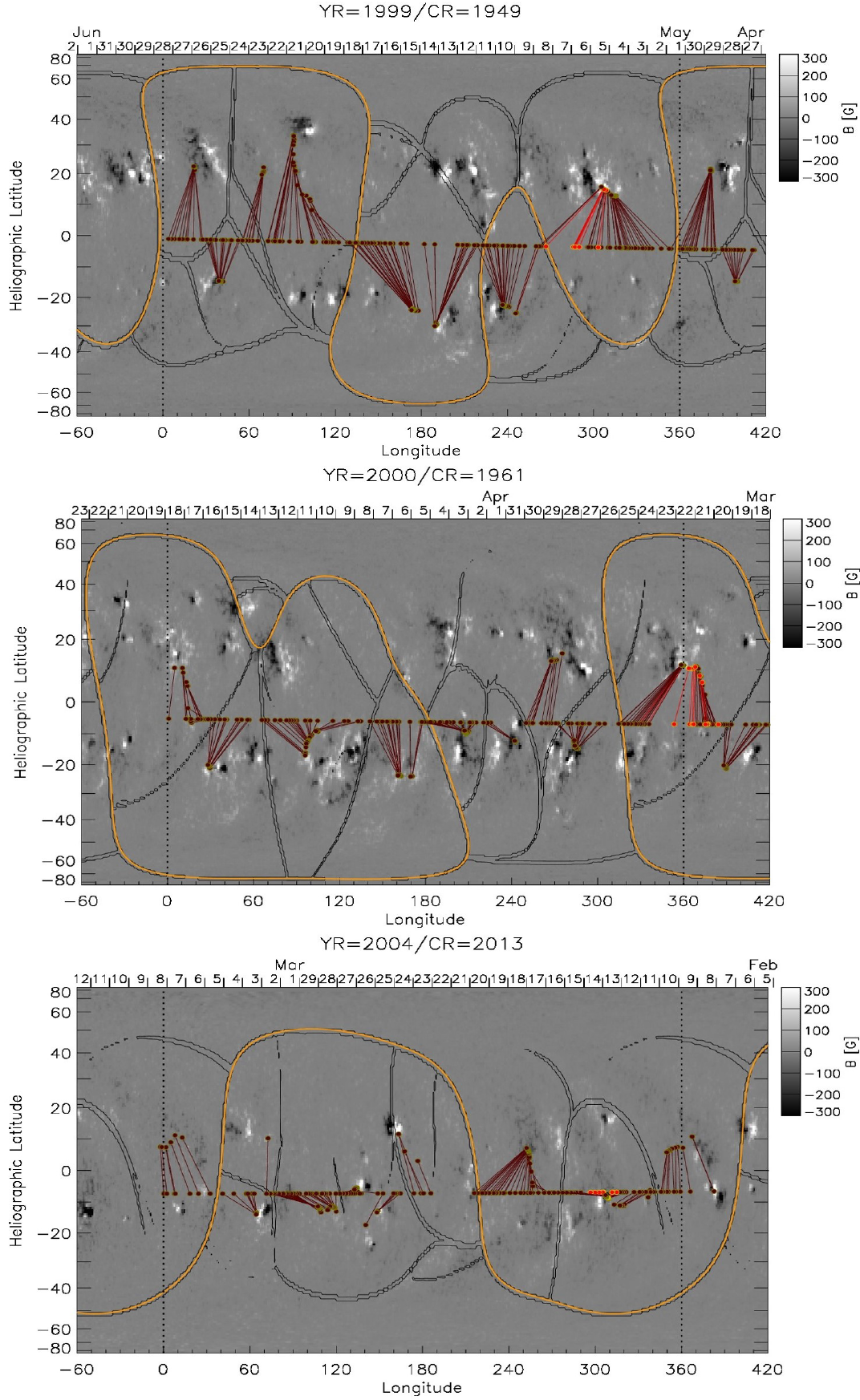}
\bigskip
\begin{minipage}{12cm}
\caption{Synoptic maps, made using {{SOHO/MDI}} magnetograms, during CR1949 (top),  CR1961 (middle), and CR 2013 (bottom) corresponding to May 1999, Mar 2000, and Feb 2004 events. Regions of strong magnetic fields corresponding to active region locations are shown as black and white patches representing negative and positive magnetic polarities, respectively. The solid curved lines in yellow on each map are the source surface magnetic neutral lines. The groups of converging lines in red and magenta colors are potential field computed \citep{HKo99} magnetic field lines that join the source surface at 2.5 R$_{\odot}$ to their counterparts at the photosphere. The lines in red in each panel particularly indicate the footpoints of the converging field lines during the traceback period when the solar wind flow was highly non-radial.}
\label{fig3}
\end{minipage}
\end{figure}
 %----------------------------------- End Fig 3 ------------------------
\subsection{Source Region Locations}
\label{S-location}
It is seen from Fig. \ref{fig2} that the source regions of the prolonged and highly non-radial solar wind flows showed a clear transition from a CH to an AR located near the CH during the non-radial flow event period. Thus, in the first step, a velocity traceback method has been employed to find source regions of the non-radial flow events. In the traceback method, the solar wind flows during the event period are usually tracked back, using in-situ measurements of solar wind speed from the {{ACE}} spacecraft at L1 to the source surface at 2.5 R$_{\odot}$. The event period at the source surface corresponding to the event at 1 AU is the traceback date. The velocity traceback technique is usually employed for steady-state solar wind flows. However, it has also been successfully used even for the highly non-radial flow events (e.g.,  \cite{BaJ03, JaF05, JaF08, JaT08}) during the well-known solar wind disappearance event of 11 May 1999. In the second step, using potential field computations \citep{HKo99} to synoptic maps, the field lines from the source surface to the photosphere are traced back. Then, the background magnetic field conditions on the Sun during our period of interest are investigated.

Synoptic maps showing the heliographic latitudes and longitudes of strong magnetic fields during Carrington rotations (CR) 1949, 1961, and 2013 are shown {{in Fig. \ref{fig3} (top), (middle) and (bottom), respectively,}} corresponding to the events of May 1999, Mar 2000 and February 2004. The synoptic maps were made 
using magnetograms obtained from the {{Michelson-Doppler-Imager}} ({{MDI}}; \cite{ScB95}) instrument on board the {{Solar and Heliospheric Observatory}} ({{SOHO}}: \citep{DFP95}) spacecraft. In each synoptic map, the dates of the central meridian passage (CMP) are marked at the top, while the heliographic longitude ranging from 0$^{\circ}$ to 360$^{\circ}$ (indicated by two dotted vertical lines) and latitude from -90$^{\circ}$ to 90$^{\circ}$ are marked on the abscissa and ordinate, respectively. 
In Fig. \ref{fig3}, the potential field computed source surface magnetic fields lie in an equally spaced grid along the equator. On the other hand, their radially back-projected photospheric footpoints lie in tightly bunched regions associated with active regions north and south of the equator. The footpoints mark the source regions of the open magnetic field lines that connect from 1 AU to the Sun at the time of the event. The group of converging field lines shown in red and magenta in Fig. \ref{fig3} are potential field computed \citep{HKo99} magnetic field lines that join the source surface at 2.5 R$_{\odot}$ to their counterparts at the photosphere. The field lines are indicated in red to show the footpoints of converging field lines during the traceback period corresponding to the central meridian passage dates at 1 AU when the solar wind was found to be highly non-radial. The curved black line represents the pseudostreamer locations estimated at the so-called source surface at 2.5 R$_{\odot}$.

It is evident from Fig. \ref{fig3} that the back-projected photospheric footpoints, indicated by the converging field lines in red colors, during the traceback dates (05 May 1999, 21-22 March 2000, and 13 Feb 2004) for all the events are originated in or around large active regions. 
Next, daily active region maps of the solar photosphere on the event and traceback dates were examined (Figures not shown here). On the traceback date of 05 May 1999 corresponding to the event date of 11 May 1999, a large active region AR 8525 was almost precisely located at the central meridian. Similarly, on the traceback dates of 23 Mar 2000 and 13 Feb 2004, large active regions AR 8916 and AR 10554 were found to be located $\sim5^{\circ}$ east of the central meridian. It is already known that the solar wind flows originating at the Sun's central meridian are usually Earth-directed \citep{JaF05}. It is to be noted that the source ARs for all the events were located at or close to ($\sim5^{\circ}$-$10^{\circ}$) the central meridian when the solar wind flows become non-radial. It is argued \citep{JaF08} that Earth-directed flows from an AR located east of the central meridian would take more time to transit across the central meridian. Therefore, for the events of Mar 2000 and Feb 2004, the duration of non-radial flow is longer than in May 1999, where the corresponding AR was located at the central meridian.
%----------------------------- Begin Fig 4-----------------------
% \begin{center}
%  	\protect\begin{figure*}[!ht]
% 	\centering
%  		\includegraphics[width=.95\textwidth, height=.35\textheight]{Fig4.pdf}
%  		\caption{The 3D structure of coronal magnetic fields during the traceback dates for (a) 22 Mar 2000 in CR 1960 and (b) 13 Feb 2004 in CR 2013, viewed, respectively, from a central meridian passage longitude of 30$^{\circ}$ and 270$^{\circ}$. 
% }
%  			\label{fig4} 
%  	\end{figure*}
% 	\vspace{-0.1cm}
%  \end{center}
\begin{figure}
\centering
\includegraphics[scale=0.6]{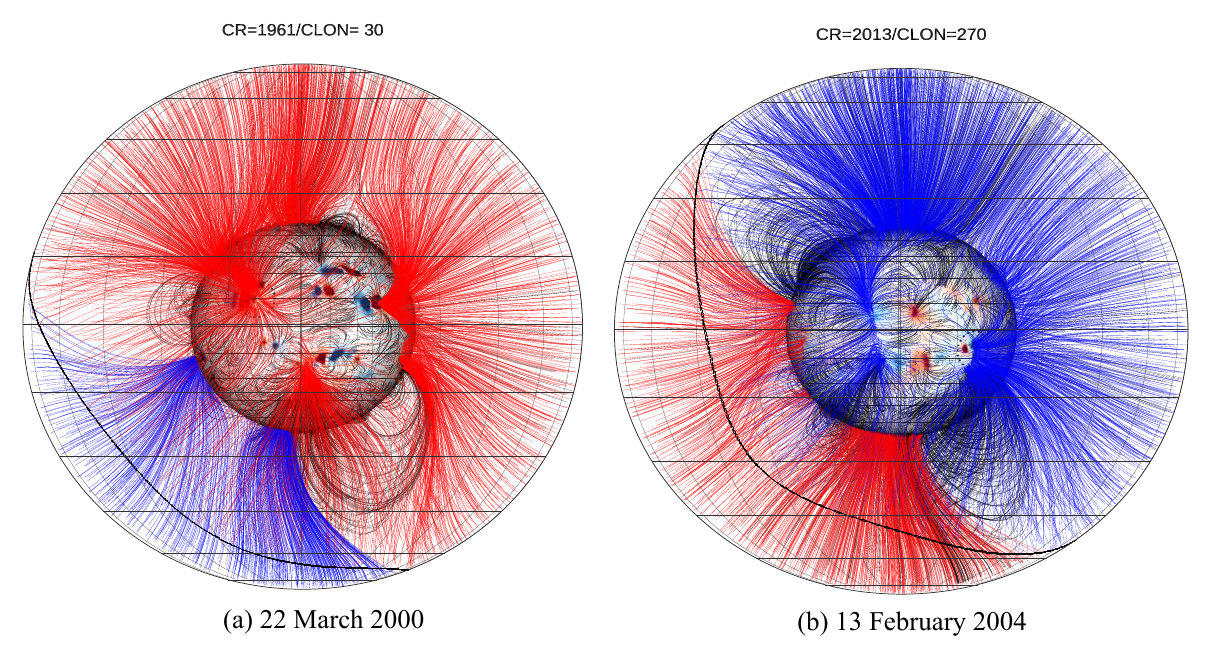}
\bigskip
\begin{minipage}{12cm}
\caption{The 3D structure of coronal magnetic fields during the traceback dates for (a) 22 Mar 2000 in CR 1960 and (b) 13 Feb 2004 in CR 2013, viewed, respectively, from a central meridian passage longitude of 30$^{\circ}$ and 270$^{\circ}$.}
\label{fig4}
\end{minipage}
\end{figure}
%----------------------------------- End Fig 4------------------------
%

The 3D coronal magnetic field configurations associated with the AR during the traceback dates of 22 Mar 2000 in CR 1960 and 13 Feb 2004 in CR 2013, viewed, respectively, from a CMP longitude of 30$^{\circ}$ and 270$^{\circ}$ are shown in Fig. \ref{fig4}. The magnetic field lines were obtained using a potential field source surface (PFSS) model \citep{HKo99} and projected to a source surface at 2.5 R$_{\odot}$. The field lines in red and blue represent opposite polarities, with red being the outward-going or positive fields and blue being the inward-going or negative fields. The thin lines in black are closed magnetic field loops; the thick black curve is the source surface magnetic neutral line. Only field lines with magnitudes of 5--250 G are shown.
Similarly, the 3D coronal magnetic field structure on the traceback date of 05 May 1999 for the May 1999 non-radial event can be referred to in \cite{JaF05}. The 3D coronal magnetic field configurations associated with the source AR of the respective events on the traceback dates clearly show open magnetic fields. The outward-going open field lines in red for the Mar 2000 event (see Fig. \ref{fig4} (left)) and inward-going open field lines in blue for the Feb 2004 event (see Fig. \ref{fig4} (right)) are observable associated with the AR. To better depict the uncluttered open field lines associated with the source AR of the non-radial flow events, the AR locations of the respective events shown in Fig.\ref{fig4} are not precisely located at or close to the central meridian. 
%
%----------------------------- Begin Fig 5-----------------------
% \begin{center}
%  	\protect\begin{figure*}[!htb]
% 	\centering
%  		\includegraphics[width=.85\textwidth, height=.65\textheight]{Fig5.pdf}
%  		\caption{An image cutout of the AR--CH complex region associated with the 11 May 1999 event. The images on the left and the right are, respectively, on the traceback date of 05 May 1999 and the next day, 06 May 1999. The left and right panels, from top to bottom, show the images of {{EIT}} 171  {\AA}, {{EIT}} 195 {\AA}, {{EIT}} 284 {\AA}, and {{SOHO}}/{{MDI}}, respectively. The white arrows in the top panels indicate the new CH features, while the black arrows mark the emergence of new magnetic fluxes. }
%  			\label{fig5} 
%  	\end{figure*}
% 	\vspace{-0.1cm}
%  \end{center}
\begin{figure}[!ht]
\centering
\includegraphics[width=.75\textwidth, height=.65\textheight]{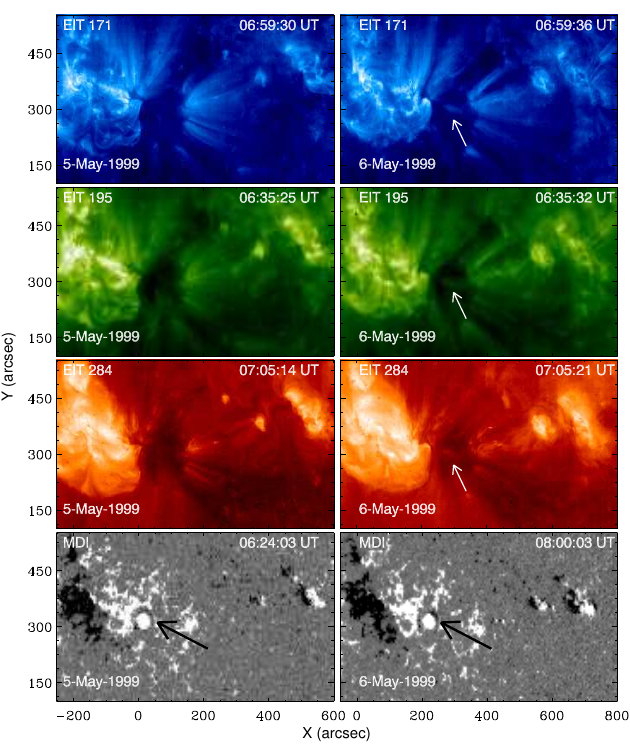}
\bigskip
\begin{minipage}{12cm}
\caption{An image cutout of the AR--CH complex region associated with the 11 May 1999 event. The images on the left and the right are, respectively, on the traceback date of 05 May 1999 and the next day, 06 May 1999. The left and right panels, from top to bottom, show the images of {{EIT}} 171  {\AA}, {{EIT}} 195 {\AA}, {{EIT}} 284 {\AA}, and {{SOHO}}/{{MDI}}, respectively. The white arrows in the top panels indicate the new CH features, while the black arrows mark the emergence of new magnetic fluxes.}
\label{fig5}
\end{minipage}
\end{figure}
%----------------------------------- End Fig 5------------------------
\subsection{Source Region Evolutions}
\label{S-evoln}
It is clear from Fig. \ref{fig3} and Fig. \ref{fig4} that the observed non-radial flow events at 1 AU are associated with large photospheric active regions and open field regions in the corona on the traceback periods. Also, it is observed from Fig. \ref{fig2} that the non-radial flow source regions have shown a transition from CH to AR. To investigate how the source regions of the non-radial flow events showed a transition from CH to AR, the {{Extreme ultra-violet Imaging Telescope}} ({{EIT}}; \cite{DeA95}) images at different wavelengths and {{MDI}} maps on board {{SOHO}} spacecraft are plotted. In Figure \ref{fig5}, the panels on the left and right, starting from the top to bottom, show cut-off images of {{EIT}} 171 {\AA}, {{EIT}} 195 {\AA}, {{EIT}} 284 {\AA}, and {{SOHO}}/{{MDI}}, respectively, for the event of May 1999. The images on the left are on the traceback date of 05 May 1999, while those on the right are 24-hour intervals after, {\it{i.e.,}} on 06 May 1999.

Similarly, Figures showing the same for Mar 2000 and Feb 2004 (not shown here) are prepared. The cut-off images of {{EIT}} for all the events show AR-CH complexes located at the central meridian. The corresponding photospheric changes of the evolving CH region for the 1999 May event are shown in the bottom panel of Figure 5. The magnetic field strength of the region displayed in the magnetogram has been limited to $\pm$300 G. The image in the left panel is at 06:24:03 UT on 05 May 1999, while the image in the right panel is a little over 24 hours later at 07:59:02 UT on 06 May 1999. Each panel's black and white regions correspond to negative and positive polarities. Corresponding to the location of the AR complex, as seen in the corona from the {{EIT}} image, is a small, white, circular region of strong magnetic fields located at the central meridian, which is part of a large bipolar magnetic region. On 06 May, a new negative polarity, indicated by a black arrow, appears northwest of the circular sunspot region. Correspondingly, the two bipolar regions located to the far west also appear to evolve with new magnetic flux elements emerging. Similar to the May 1999 event, the corresponding photospheric changes of the evolving CH region for the Mar 2000 and Feb 2004 events (Figures not shown here) are noticed. It is found that the Mar 2000 and Feb 2004 events also saw the emergence of new magnetic flux elements around the source region of interest.

\begin{figure}
\centering
\includegraphics[width=.65\textwidth, height=.35\textheight]{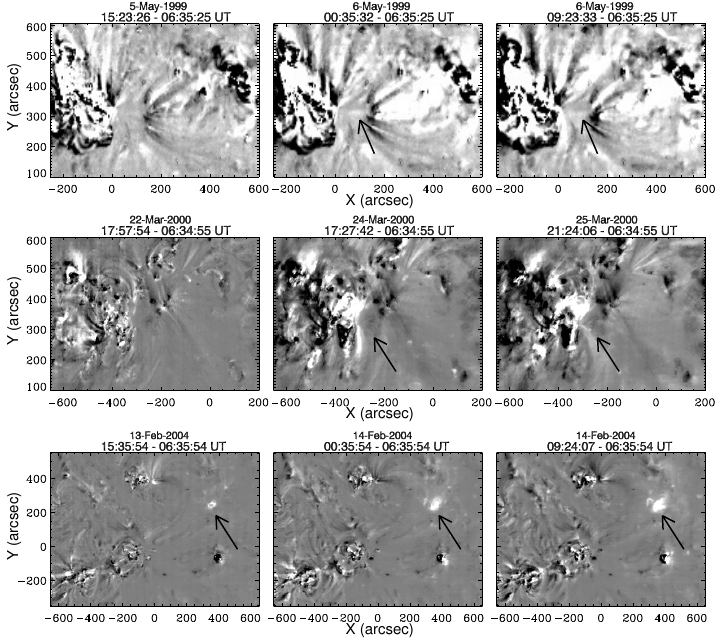}
\bigskip
\begin{minipage}{12cm}
\caption{Base difference images were obtained at an interval of 9 hours from {\it{EIT}} images corresponding to May 1999, Mar 2000, and Feb 2004 events. The arrows in the middle and right side panels mark the changes in the CH region.}
\label{fig6}
\end{minipage}
\end{figure}

Next, base difference images of the Sun from the {{EIT}} are obtained, as shown in Fig. \ref{fig6} during the traceback periods of 5\,--\,8 May 1999, 21\,--\,23 Mar 2000 and 13\,--\,15 Feb 2004.  The difference images are prepared by subtracting an image on the traceback date (called a reference image) from an image $\sim$9 hours ahead of the reference image. There is no specific reason for using a 9-hour difference between the reference and base images. The evolutions in the AR-CH complexes are then investigated using the base difference images thus produced. The base difference images, prepared at a 9-hour difference, clearly show the evolutions of the features in the AR-CH complexes. Therefore, the 9-hour difference images are shown in Fig. \ref{fig6}.
The black regions in the base difference image are treated as regions of the reference image, while the white regions are considered new features. From Fig. \ref{fig6}, it is seen that CH regions are located close to or at the central meridian on the traceback dates. The very next day, new bright features emerged at the center of the CH region, as indicated by black arrows in the right-hand panels of Fig. \ref{fig6}. The new features observed in the base difference images ahead of 9 hours from the reference image are perceived as constrictions or narrowing of the CH. Generally, the CH are regions of the fast wind streams. With the emergence of the bright features in the CH, the area of the CH is further reduced. There is a high degree of correlation between solar wind speed and the size of the CH from which it emanates (Kojima et al. 1999). As a result, the flow speed of the stream originating from the CH is decreased further, and thus, the appearance of the bright features in the CH results in the emergence of the slow wind streams from the CH.
%----------------------------- Begin Fig 6-----------------------
% \begin{center}
%  	\protect\begin{figure*}[!ht]
% 	\centering
%  		\includegraphics[width=.7\textwidth, height=.5\textheight]{Fig6.pdf}
%  		\caption{Base difference images were obtained at an interval of 9 hours from {\it{EIT}} images corresponding to May 1999, Mar 2000, and Feb 2004 events. 
% 		The arrows in the middle and right side panels mark the changes in the CH region.}
%  			\label{fig6} 
%  	\end{figure*}
% 	\vspace{-0.1cm}
%  \end{center}
%----------------------------------- End Fig 6 ------------------------
%
\section{Discussion and Conclusion} 
The present study reports three extremely non-radial solar wind outflow events on 10--12 May 1999, 27--30 Mar 2004, and 15--18 Feb 2004. In all events, the average solar wind flow deviations from the radial direction exceed 6$^{\circ}$ for a prolonged period between 24--48 hours or more. Generally, prolonged and extremely non-radial flows have been observed during CMEs and CIRs. However, none of them were associated with CMEs and CIRs. In addition to the extremely non-radial flow, for all of the events, the solar wind flow velocities and densities were lower than usual during the duration of the non-radial flow. Such low-density and low-velocity solar wind flow has been studied earlier for solar wind disappearance events \citep{BaJ03, JaF05, JaF08, JaT08}.

The Oxygen charge state ratio at L1 for all of the events has increased from 0.2 to $>$0.2 just when the angle of radial solar wind flow exceeds 6$^{\circ}$, {\it{i.e.}}, the flow becomes highly non-radial. This observation shows that the source regions must originate from AR-CH complexes so that the source of the solar wind flow can make a smooth transition from CH to AR \citep{ScC03,JaF08}. 
Further, the solar wind flows have been tracked from 1 AU to the source surface at 2.5 R$_{\odot}$ using a constant velocity traceback method. Next, using potential field computations applied to a synoptic magnetogram, the footpoints of the magnetic field lines from the source surface to the photosphere are traced back. It is found that all of the events on the traceback dates of the events are found to be associated with large AR. The daily maps of the magnetogram on the corresponding traceback dates show that these source ARs are located at or close to the central meridian. In addition, the three-dimensional magnetic field configurations of the Sun obtained using PFSS on the traceback date indicate the presence of open field regions next to the ARs.
On the other hand, the {{EIT}} cut-off images as well as the {{EIT}} base difference images on the traceback dates of the events, have shown small, mid-latitude CH adjacent to the large AR. Further, it is revealed from {{EIT}} images that the CH evolves in time and shows the emergence of bright coronal loops. Similarly, the {{MDI}} cut-off images on the traceback dates of the events show the emergence of opposite polarity magnetic flux elements on the photosphere close to the source ARs. 
The emergence of bright coronal loops progressively reduces the CH area. It is found that there is a high degree of correlation between solar wind speed and the size of the CH from which it emanates \citep{KoF99}. The reduction in the size of the CH would, therefore, suppress the CH outflow, giving rise eventually to a low-velocity solar wind stream. The rapid evolution observed in the CH complex before the start of the non-radial flow has been attributed to interchange reconnections. The reconnection occurs between the open unipolar CH fields and the closed bipolar magnetic flux regions \citep{JaT08}.

Based on our observations, we propose a possible explanation for the cause of the prolonged and highly non-radial flows. Generally, in a typical CIR, a compression region is formed at the leading edge of the faster solar wind stream emanating from a CH when the faster solar wind stream overtakes the slower solar wind stream adjacent to the faster stream. In contrast, in our case, because of the reduction in the size of the CH, the faster solar wind stream becomes even slower than the average slow solar wind flowing adjacent to it. As a result, a compression region is formed at the trailing edge of the significantly slower stream. As suggested by \citet{Piz89, Piz91}, the stream interface of the CIRs acts like a slowly moving wall in the interplanetary medium. The compression region, thus formed, acts as a slowly moving wall.  Subsequently, as time progresses, the source CH further reduces in size and finally vanishes, leaving only the AR behind. This results in the slowly moving wall being ``pinched off" into the interplanetary space. The local interplanetary magnetic field, thus, gets diverted by the slowly moving wall \citep{LOM19} because of the prevailing low beta plasma condition with the magnetic field strength during the events close to average and the very low solar wind density. The incoming solar wind flows follow the deflected magnetic field, possibly leading to extremely non-radial solar wind flow. Therefore, studying such prolonged and extremely non-radial solar wind events is crucial to help us gain further insights into the highly non-radial flow not associated with CMEs/CIRs.

\begin{acknowledgments}
This work has made use of NASA's OmniWeb services Data System. The authors thank the free data use policy of the National Solar Observatory (NSO). JP and DR acknowledge the ISEE International Collaborative Research Program for support in executing this work. We thank the anonymous reviewer for his/her constructive and insightful comments that helped us substantially improve the paper.

\end{acknowledgments}

\begin{furtherinformation}
%\begin{orcids}
%\orcid{0000-0002-9448-1794}{Susanta Kumar}{Bisoi}
%\orcid{0000-0002-7820-7971}{Diptiranjan}{Rout}
%\orcid{0000-0003-2504-2576}{P.}{Janardhan}
%\orcid{0000-0002-7060-3750}{K.}{Fujiki}
%\end{orcids}
\begin{authorcontributions}
DR, PJ, and DC designed the project. DR, SKB, and PJ wrote and edited the manuscript. DR, KF, SKB, and KS performed the analyses and prepared the figures. DR, SKB, PJ, KF, and DC discussed the results and implications of the work at all stages.
\end{authorcontributions}

\begin{conflictsofinterest}
The authors declare no conflict of interest.
\end{conflictsofinterest}

\end{furtherinformation}

\bibliographystyle{bullsrsl-en}

\bibliography{extra}

\end{document}